%% file: ms.tex
\documentclass{article}

\usepackage{amsmath}
\usepackage{multirow}
\usepackage{amsthm}
\usepackage{bbm}

\usepackage{amssymb}

\usepackage[export]{adjustbox}

\DeclareMathOperator{\Sym}{Sym}
\DeclareMathOperator{\Pos}{Pos}

\DeclareMathOperator{\LC}{LC}

\newtheorem{proposition}{Proposition}
\newtheorem{theorem}{Theorem}
\newtheorem{lemma}{Lemma}

\theoremstyle{definition}
\newtheorem{definition}{Definition}
\newtheorem{example}{Example}

\usepackage[preprint]{neurips_2021}

\usepackage[utf8]{inputenc} 
\usepackage[T1]{fontenc}    
\usepackage{hyperref}       
\usepackage{url}            
\usepackage{booktabs}       
\usepackage{amsfonts}       
\usepackage{nicefrac}       
\usepackage{microtype}      
\usepackage{xcolor}         
\usepackage{backref}

\usepackage{xspace}

\usepackage{bm}
\usepackage{float}
\usepackage{natbib}
\usepackage{graphicx}

\definecolor{ao(english)}{rgb}{0.0, 0.5, 0.0}

\newcommand{\name}{\textit{NeuraCrypt}\xspace}

\title{\name: Hiding Private Health Data\\ via Random Neural Networks \\for Public Training}

\author{%

  Adam~Yala$^{\dagger}$\thanks{Equal contribution}~, Homa~Esfahanizadeh$^{\dagger*}$, Rafael~G.~L. D’Oliveira$^{\dagger}$, Ken ~R.~Duffy$^\ddagger$,\AND Manya Ghobadi$^\dagger$\vspace{-1cm}, Tommi~S.~Jaakkola$^\dagger$,  Vinod~Vaikuntanathan$^\dagger$ \AND Regina Barzilay$^{\dagger}$, Muriel M\'{e}dard$^{\dagger}$\AND\vspace{-0.0cm}\\
  $\dagger$Massachusetts Institute of Technology (MIT), 02139 USA \\
  $\ddagger$Maynooth University, Ireland\\
      \texttt{\{adamyala, ghobadi, tommi, regina\}@csail.mit.edu}
  \\
  \texttt{\{homaesf, rafaeld, vinodv, medard\}@mit.edu}  \\
  \texttt{Ken.Duffy@mu.ie} \\
}

\begin{document}

\maketitle

\begin{abstract}

\input{abstract}

\end{abstract}

\section{Introduction}
\input{intro}

\section{Related Work}
\input{related_work}

\section{Method} \label{sec: formulation}

\input{method} 

\section{Privacy Analysis} \label{sec:theory}
\input{security}

\section{Experiments}\label{sec:implementation}
\input{experiments}

\section{Results}
\input{results}

\section{Conclusion}
\input{discussion}

\bibliographystyle{plainnat}
\bibliography{references.bib}




\appendix
\section{Broader impact}
\input{impact}

\section{Proofs of theoretical analysis}
\input{appendixA}

\section{Additional experiments}
\input{addit_experiments}

\end{document}

%% file: abstract.tex
Balancing the needs of data privacy and predictive utility is a central challenge for machine learning in healthcare. In particular, privacy concerns have led to a dearth of public datasets, complicated the construction of multi-hospital cohorts and limited the utilization of external machine learning resources. To remedy this, new methods are required to enable data owners, such as hospitals, to share their datasets publicly, while preserving both patient privacy and modeling utility. We propose 
\name\footnote{github.com/yala/NeuraCrypt}, a private encoding scheme based on random deep neural networks. \name encodes raw patient data using a randomly constructed neural network known only to the data-owner, and publishes both the encoded data and associated labels publicly. From a theoretical perspective, we demonstrate that sampling from a sufficiently rich family of encoding functions offers a well-defined and meaningful notion of privacy against a computationally unbounded adversary with full knowledge of the underlying data-distribution. We propose to approximate this family of encoding functions through random deep neural networks. Empirically, we demonstrate the robustness of our encoding to a suite of adversarial attacks and show that \name achieves competitive accuracy to non-private baselines on a variety of x-ray tasks. Moreover, we demonstrate that multiple hospitals, using independent private encoders, can collaborate to train improved x-ray models. Finally, we release a challenge dataset\footnote{github.com/yala/NeuraCrypt-Challenge} to encourage the development of new attacks on \name.

%% file: intro.tex
One of the central challenges of developing machine learning tools in healthcare is access to patient data. To protect patients' privacy, regulations such as HIPPA~\citep{HIPAA} and GDPR~\citep{GDPR} greatly complicate creation of large, multi-institutional datasets, a necessary resource for training robust and equitable models. Consequently, the lack of public datasets has excluded the broad machine learning research community from contributing to clinical AI. Addressing this challenge is the focus of our paper. Consider the scenario where a hospital wishes to publicly release a dataset of mammograms with cancer labels. We are interested in developing a computationally effective mechanism for image encoding that both protects patient privacy (i.e., hiding the correspondence between patients and encoded samples), and facilitates learning. 

There are a number of solutions that have been developed for different notions of privacy. While federated learning \citep{mcmahan2017communication} can offer privacy by training models in a distributed fashion, the framework relies on the coordination of data owners and model developers to run shared software. This framework is not designed to enable hospitals to deposit their datasets publicly. Cryptographic methods, such as secure multi-party computation, fully homomorphic encryption and functional encryption \citep{Gentry09,BV11,BSW12,cho2018secure} could enable such public sharing and offer extremely strong security guarantees, hiding everything about the data. However, these security guarantees come at the cost of high computational overheads for use with modern  methods. The notion of security adopted by these cryptographic methods is not suited to our setting,  where image labels are public. Instead, we seek an efficient encoding scheme to protect the information not already implied by the image label.

We propose \name, a private encoding scheme designed to enable data owners to publish their data publicly while preserving both data privacy and modeling utility. \name encodes raw patient data using a randomly constructed neural network, known only to the data owner, and deposits both the encoded  data and associated labels publicly for unknown third parties to develop models. When applied in the multi-institutional setting, each site utilizes independent private encoders to encrypt their data. With the help of label information, models trained in this setting can map these independently constructed encodings into a shared feature space. While the data remains private across the sites, each institution can still benefit from the larger combined dataset.

Our design is guided by theoretical considerations. First, we demonstrate a means, albeit an inefficient one, of offering for each sample anonymity amongst all other possible samples sharing the same label. Our approach relies on random selection of an encoder from a sufficiently rich family of encoding functions. Secondly, while we cannot construct the  optimal family of encoding functions directly, i.e. all possible bijections, we demonstrate that we can iteratively enrich encoder families through  function composition.  These results, as well as recent work ~\citep{DGKP20} on the properties of random deep neural networks, motivate us to approximate the optimal family of encoding functions as deep neural networks. While our theoretical results do not guarantee the security of our specific \name network architecture, we provide empirical experiments to test the robustness of our approach against modern attacks, following standard practice in cryptanalysis~\citep{standard2001announcing, dworkin2015sha}. 

We empirically test our method on two benchmark chest x-ray datasets MIMIC-CXR \citep{johnson2019mimic} and CheXpert \citep{irvin2019chexpert}, and compare it against the model operating on raw data. We demonstrate that, across a variety of diagnostic tasks, \name-based models achieve competitive performance. Moreover, we show that combining multiple datasets, using separate private encoders, enables the model to benefit from additional training data and thereby improve its accuracy. \name appears robust to adversarial attacks designed to either uncover the private encoder or to uncover additional sensitive attributes from the encoded data.

We believe our \name architecture, as well as the core idea of using random deep neural network encodings to achieve privacy, provide a novel direction for privacy-preserving machine learning. We encourage the development of new attacks on \name as well as further refinements of our method, and to this end, we release a challenge dataset.

%% file: related_work.tex
\paragraph{Cryptographic techniques.} Cryptographic techniques, such as secure multiparty computation and fully homomorphic encryption~\citep{Yao86,GMW87,BGW88,DBLP:conf/stoc/ChaumCD88,Gentry09,BV14,cho2018secure}, allow one or more data owners, such as hospitals, to encode (encrypt) their data before providing them to a third party,  say a cloud data center, for computation. Building models with homomorphic encryption requires leveraging specialized cryptographic primitives, an approach that requires an impractical overhead for modern deep learning models. In contrast, \name encodings can be directly leveraged by standard deep learning techniques.

The high complexity of constructing and running homomorphic encryption generally provide extremely strong guarantees, such as a semantic security~\citep{GM82}, wherein no information regarding the original data may be leaked by the encoding or computation. However, this strong security guarantee is an overkill for our context. For instance, we do not seek to hide the fact that a hospital hosts chest x-rays or the disease labels of those x-rays, as we release the labels publicly. Our goal is to avoid expending design and run-time resources~\citep{MohasselZ17,MiniONN,GAZELLE,crypto-2018-28796} on hiding these facts through homomorphic encryption. Rather, we seek to provide privacy by protecting the characteristics of a chest x-ray not already captured by the disease label. 

\paragraph{Federated learning and differential privacy techniques.} Federated learning (FL) \citep{mcmahan2017communication} enables collaborative learning among different data-owners (hospitals) through distributed training.  The core idea of FL is to avoid transferring raw data by allocating an instance of the ML model at each data owner, and instead sharing model updates \citep{RiekeNature2020}. Despite a considerable amount of research in this area, including progress in secure aggregation \citep{bonawitz2017practical} and differential privacy \citep{dwork2014algorithmic}, preserving privacy while maintaining modeling accuracy in FL remains an open challenge \citep{FLoC, mcmahan2021advances}. In contrast to \name, FL is not designed to enable data-owners to publicly deposit their data.

\paragraph{Lightweight encoding techniques.} Our approach is most closely related to prior focused on achieving privacy through lightweight encoding schemes. ~\citep{Ko2020IeeeAccess,TanakaICCE2018,SirichotedumrongICIP2019} have proposed de-identification techniques to carefully distort images to reduce their recognition rate by humans while preserving the accuracy of image classification models. Unfortunately, such methods do not offer privacy against realistic attacks. Recently, InstaHide ~\citep{instahide} proposed to encode images by linearly mixing them with other samples and applying a pixel-wise mask. While this approach provides collaborative learning for multiple data owners, its linear transforms preserve the relative distance between two samples in the original domain and the encoded domain. This drawback was exploited by \citep{carlini2021private} to decrypt the Instahide dataset challenge. DAUnTLeSS~\citep{DAUnTLeSS} proposed to encode images using fully-connected neural networks and linear sample mixing, and analyzed the computational difficulty of reversing this encoding to an attacker with access to parallel raw and encoded data pairs. They demonstrate that this encoding is easy to reverse if the source data distribution has low entropy (e.g. MNIST), and more difficult for complex datasets. Moreover, they demonstrate that mixing samples can increase the difficulty of the recovery task. In contrast to DAUnTLeSS, we consider a threat model where the adversary is computationally-unbounded and does not have access to parallel data.

%% file: method.tex
\begin{figure}
    \centering
    \includegraphics[width=0.55\textwidth]{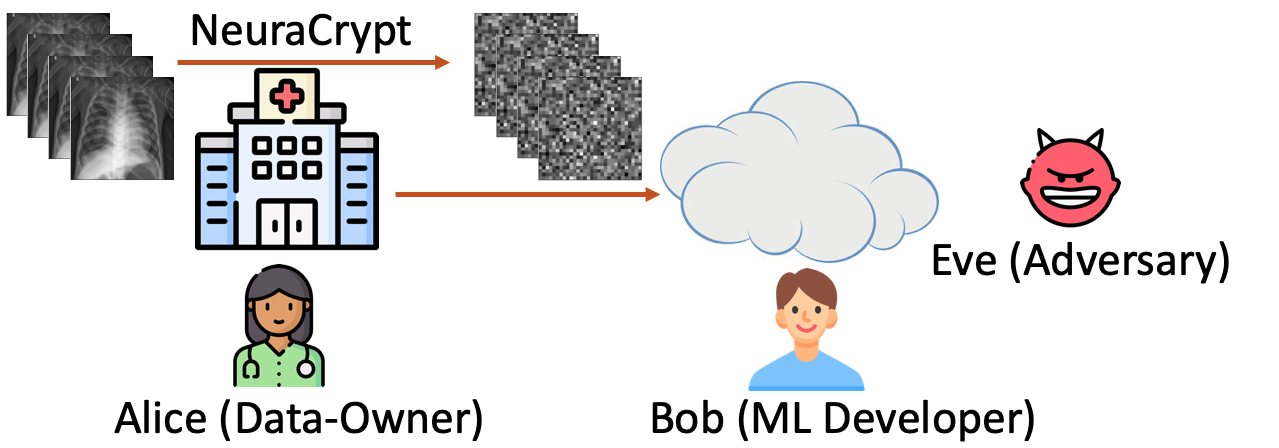}
    \caption{Alice (data-owner) transmits her labeled encoded data to Bob (ML developer). Eve (adversary) attempts to identify information about Alice's raw data beyond their labels.}
    \label{fig:alice-bob-eve}
\end{figure}

\textbf{\name.} The problem setting is depicted in Figure \ref{fig:alice-bob-eve}. We wish to enable a hospital to publish their image dataset $X_d$ with diagnostic labels $Y_d$ while protecting patient privacy.  Given the dataset $\{(x,y)\}_{x \in X_d}$, where $y=L(x)$ is the label, a data-owner randomly samples a private \name encoder $T_d$, a random neural network, and uses $T_d$ to produce encoded samples $Z_d=T_d(X_d)$, i.e., $z_i=T_d(x_i)$ for every $x_i\in X_d$. The data-owner can then deposit $\{(T_d(x_i),L(x_i))\}_{x_i\in X_d}$ publicly for untrusted third parties to develop models to estimate $\Pr[\bm{Y_d}=Y_d|\bm{Z_d}=Z_d]$. We note that multiple data owners can seamlessly collaborate to develop joint models by publishing datasets on the same task while using independent \name encoders. Given that model developers can only estimate $\Pr[\bm{Y_d}=Y_d|\bm{Z_d}=Z_d]$ and not $\Pr[\bm{Y_d}=Y_d|\bm{X_d}=X_d]$, only data-owners can directly utilize the learned models, creating an incentive for data-owners and model developers to collaborate for model dissemination. 

\begin{figure}
    \centering
    \includegraphics[width=0.99\textwidth]{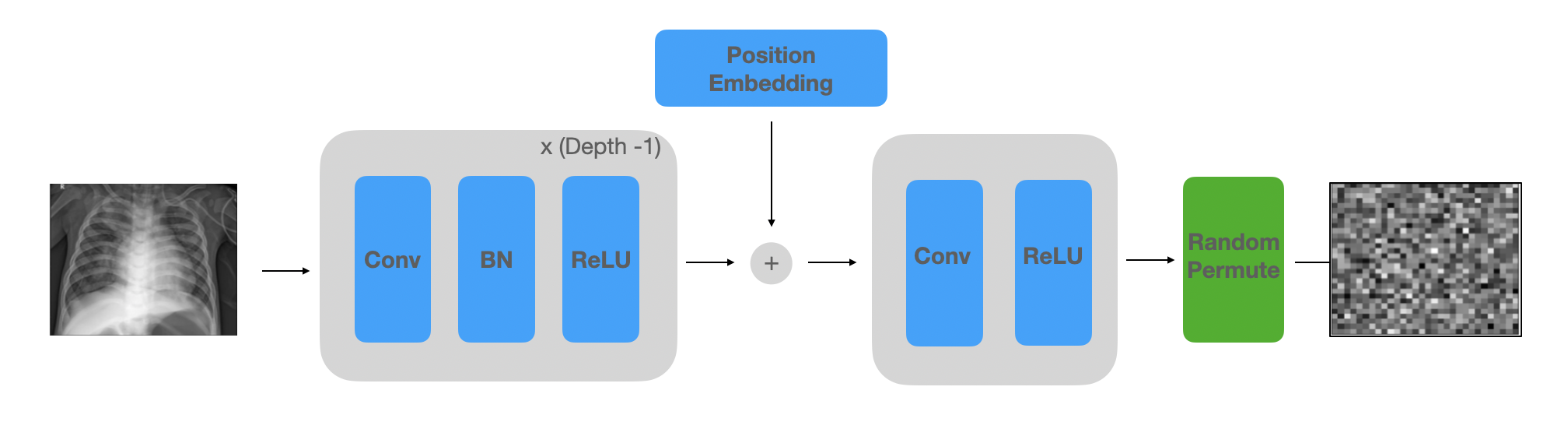}
    \caption{Architecture of \name encoder.}
    \label{fig:architecture}
\end{figure}

While many \name architectures are possible, we focus on medical imaging tasks, and thus implement our \name encoders as convolutional neural networks. Our encoder architecture is illustrated in Figure~\ref{fig:architecture}, and consists of convolutional layers with non-overlapping strides, batch normalization \citep{ioffe2015batch}, and ReLU non-linearities. To encode positional information into the feature space while hiding spatial structure, we add a random positional embedding for each patch before the final convolutional and ReLU layers and randomly permute the patches at the output independently for each private sample. This results in an unordered set of patch feature vectors for each image. We note that this architecture is closely inspired by the design of patch-embedding modules in Vision Transformer networks \citep{dosovitskiy2020image,zhou2021deepvit}.

\textbf{Threat model.} We assume a computationally unbounded adversary which knows all possible images $\mathcal{X}$ as well as their labels, i.e., $\{(x,L(x))\}_{x\in\ \mathcal{X}}$. The adversary also knows the distributions of both the data-owners' samples $\Pr [\bm{X_d}=X_d]$ and the choice of encoder function $\Pr [\bm{T_d} = T]$. This adversary knows $\{(T_d(x),L(x))\}_{x\in\ X_d}$, as the  data owner deposits it publicly. Moreover, since we do not know the adversary's classifying capabilities, we assume the worst case, in which she is able to perfectly classify the encoded samples. Thus, we assume the adversary also knows $\{(T_d(x),L(x))\}_{x\in\ \mathcal{X}}$, where now the images are taken over the whole of $\mathcal{X}$. The goal of the 
adversary is to learn more about the random variable $\bm{X_d}$ using $Z_d, Y_d$ than  from $Y_d$ alone. We define this formally in section \ref{sec:theory}.

\textbf{Privacy intuition.} We provide a formal privacy analysis for the described threat model given an encoder space $\mathcal{F}$ in Section \ref{sec:theory}. We consider a theoretical $\mathcal{F}$ which consists of all bijections from $\mathcal{X}$ to $\mathcal{X}$ with the same label assignment. These encodings map each image of the hospital's dataset to another image in $\mathcal{X}$ with the same label. This function family is exponentially large in size of $\mathcal{X}$ and a computationally unbounded adversary, as described in our threat model, cannot distinguish between its members. In this setting, observing $(Z_d, Y_d)$ does not offer the adversary more information about the underlying $\bm{X_d}$ than observing $Y_d$. While sampling from this optimal family directly is not feasible, as it would require full knowledge of $\mathcal{X}$ and its labels, our theoretical analysis demonstrates that we can enrich the privacy of our encoding scheme via functional composition. This result motivates us to approximate the $\mathcal{F}$ using deep neural networks.

%% file: security.tex
In this section, we present a detailed analysis of the threat model in Section \ref{sec: formulation}. All proofs appear in the appendix. We denote the set of all samples by $\mathcal{X}$ and assume it is a finite set. Each sample $x \in \mathcal{X}$ is labeled by a function $L: \mathcal{X} \rightarrow \mathcal{Y}$, where the set of labels $\mathcal{Y}$ is also finite.  We denote the set of all bijections from $\mathcal{X}$ to itself by $\Sym (\mathcal{X}) = \{T:\mathcal{X}\rightarrow \mathcal{X}: \text{$T$ is a bijection} \}$. As each data-owner acts independently, we perform our analysis individually. We consider a model with three participants, Alice, Bob and Eve,   to be  consistent with the common terminology of privacy  (see also Figure~\ref{fig:alice-bob-eve}): 

\noindent
\textbf{Alice (the data owner).} Alice has a private subset of samples $X_A \subseteq \mathcal{X}$ which is drawn from a distribution $\Pr [\bm{X_A}=X_A]$. The subscript $A$ for Alice replaces the generic subscript $d$ used for $X$ in our Method discussion. Alice samples an encoder $T_A$ from a family $\mathcal{F} \subseteq \Sym (\mathcal{X})$ according to a distribution $\Pr [\bm{T_A} = T]$, and then transmits $M_A(X_A) = \{(T_A(x),L(x))\}_{x\in\ X_A}$ to Bob. Alice does not know or have control over $\Pr [\bm{X_A}=X_A]$ but gets to choose $\mathcal{F}$ and the distribution $\Pr [\bm{T_A} = T]$.
    
\noindent
\textbf{Bob (the model developer).} Bob, who is not aware of the true labeling $L \in \mathcal{Y}^\mathcal{X}$, knows a prior distribution $\Pr[\bm{L}=L']$, and also the distribution $\Pr [\bm{T_A} = T]$. After receiving $M_A(X_A) = \{(T_A(x),L(x))\}_{x\in\ X_A}$, he is interested in learning a classifier on Alice's encoded data, $L_A = L \circ T^{-1}_A$, i.e., the distribution $\Pr[\bm{L_A}=L_A \mid \bm{M_A(X_A)}=M_A(X_A)]$, for every $x \in \mathcal{X}$.
    
\noindent
\textbf{Eve (the adversary).} Eve knows $\{(x,L(x))\}_{x\in\ \mathcal{X}}$ and the distributions $\Pr [\bm{X_A}=X_A]$ and $\Pr [\bm{T_A} = T]$. Since Alice (e.g a hospital) releases the data publicly, Eve also knows $M_A(X_A) = \{(T_A(x),L(x))\}_{x\in\ X_A}$. While impractical, we consider a worst case scenario where Eve  is able to classify perfectly any sample and thus, we also assume that she knows $M_A(\mathcal{X})=\{(T_A(x),L(x))\}_{x\in\ \mathcal{X}}$. Eve is interested in learning the random variable $\bm{X_A}$.

In our model we are interested in comparing what Eve learns of Alice's samples $\bm{X_A}$ from the observation of $M_A(X_A)$ and $M_A(\mathcal{X})$ compared to only having observed $Y_A = \{L(x)\}_{x\in\ X_A}$. In this context, we say Alice's scheme is \emph{perfectly private} if $\Pr[\bm{X_A} = X_A \mid \bm{M_A(X_A)}=M_A(X_A) ,\; \bm{M_A(\mathcal{X})}=M_A(\mathcal{X})] = \Pr[\bm{X_A} = X_A \mid \bm{Y_A}=Y_A]$. Since Alice can only choose $\mathcal{F}$ and its distribution $\Pr [\bm{T_A} = T]$, we often refer to $\mathcal{F}$ as Alice's scheme.

For notation purposes, we occasionally impose a total order $\preceq$ on $\mathcal{X}$ to represent it by a vector $(x_1,\ldots,x_{|\mathcal{X}|})$ such that $x_i \preceq x_j$ if $i\leq j$. We then represent a transformation $T$ by a vector with size $|\mathcal{X}|$ whose $i$-th element is $T(x_i)$. 

Next we define the \textit{label configuration (LC)} and the \textit{LC-anonymity list}, as these notions capture Eve's uncertainty about Alice's private samples $X_A$. 

\begin{definition}[Label configuration] The original label configuration, denoted by $\LC(\mathcal{X})$, is a vector of size $|\mathcal{X}|$ whose $i$-th elements is $L(x_i)$. The label configuration of an encoder $T$, denoted by $\LC(T)$, is a permutation of $\LC(\mathcal{X})$ according to $T$, i.e. the $i$-th element of $\LC(T)$ is $L \circ T^{-1} (x_i)$. 
\end{definition}

\begin{definition}[LC-anonymity list] Given a family of encoders $\mathcal{F}$, the LC-anonymity list of an encoder $T \in \mathcal{F}$ is defined as $\mathcal{F}_{T} = \{T' \in \mathcal{F}: \LC(T')=\LC(T) \}.$
\end{definition}

\begin{figure}
    \centering
    \includegraphics[width=0.75\textwidth]{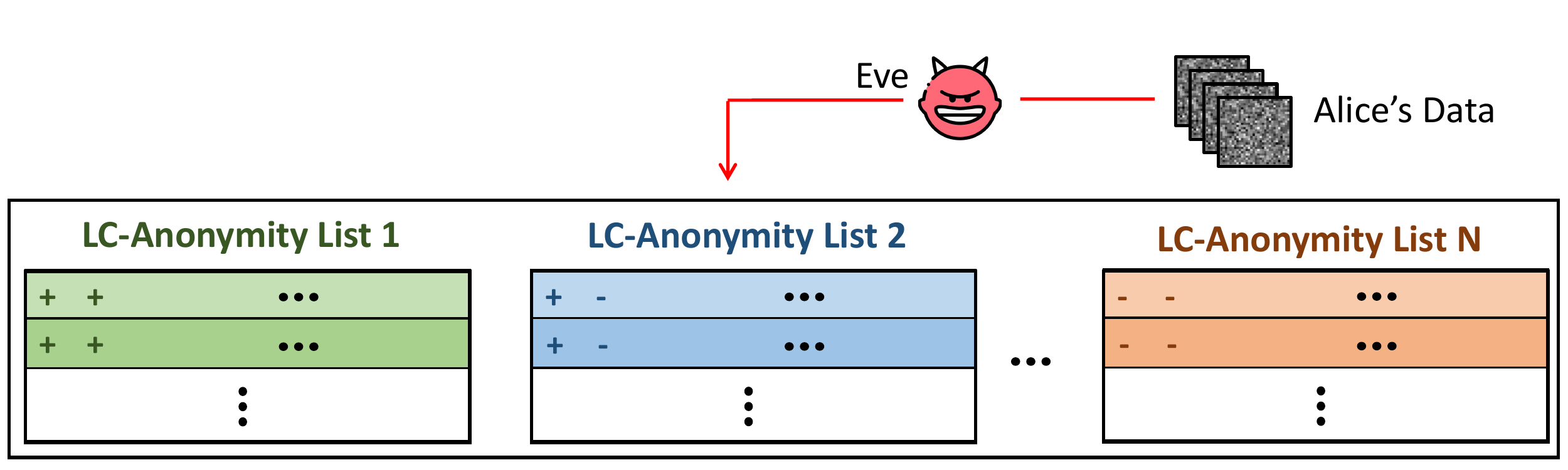}
    \caption{Eve receives Alice's encoded data and selects the correct set of  encoders $\mathcal{F}_{T_A}$ according to the observed label configuration. Eve cannot distinguish among encoders within one LC-anonymity list, and the best she can do is to identify the possible value for Alice's raw data corresponding to each encoder in the list.}
    \label{fig:EveandLCAnonimity}
\end{figure}

The LC-anonymity lists partition $\mathcal{F}$ into different equivalence classes (see Figure~\ref{fig:EveandLCAnonimity}). We also note that under these new concepts, Eve's knowledge of $M_A(X_A) = \{(T_A(x),L(x))\}_{x\in\ X_A}$ and $M_A(\mathcal{X})=\{(T_A(x),L(x))\}_{x\in\ \mathcal{X}}$ is equivalent to her knowing $T _A({X_A})$ and $\LC(T_A)$.

\begin{example} \label{sec:example1}
Let $\mathcal{X} = \{1,..., 16\}$ and, for representation purposes, consider the ordering $1\preceq 2\preceq\dots,\preceq 16$. Let $\mathcal{Y}=\{-,+\}$, and $\mathcal{F}=\Sym (\mathcal{X})$. Let the labeling $L$ be such that $\LC(\mathcal{X})=(++--++----------)$. Assume that Alice chooses $T_A=(12,2,11,4,6,8,16,15,13,7,9,5,3,14,1,10)$, 
and that her data is $X_A=\{2,5,7, 10,15\}$. Thus, she sends $\{(2,+),(6,+),(16,-),(7,-),(1,-)\}$ to Bob. According to the threat model, Eve knows $\LC(T_A)=(-+---+-+---+----)$. Although $|\mathcal{F}|=16!=2.1\mathrm{e}{13}$, since Eve knows $\LC(T_A)$, she can infer that Alice's encoder is in the smaller set $\mathcal{F}_{T_A}$ with cardinality $4!12!=1.1\mathrm{e}{10}$. Thus, if Alice had chosen $T_A$ uniformly at random from $\mathcal{F}$, then the probability of Eve guessing $T_A$ correctly would be approximately $1\mathrm{e}{-10}$. Eve, however, is interested in learning $X_A$. In this case, since $T_A$ is uniform over all permutations, it can be shown that $\mathcal{F}$ is perfectly private, i.e. Eve learns no more than she would by solely observing $Y_A = \{L(x)\}_{x\in\ X_A}$.
\end{example}

As illustrated in the example above, the scheme where Alice chooses $T_A$ uniformly from $\Sym(\mathcal{X})$ is perfectly private. This scheme, however, completely scrambles of the dataset, making Bob's learning task impossible. By observing a subset of the encoded data, Bob still cannot predict the label of a held-out sample better than the prior probability. In Theorem \ref{theorem: optimal scheme}, we show that there exist schemes which are perfectly private, while still preserving the structure of $X_A$. Towards that, we begin by investigating the privacy of general families $\mathcal{F} \subseteq \Sym(\mathcal{X})$.

We begin by characterizing the set of possible values for $X_A$, given Eve's observations.

\begin{proposition}\label{prop:possXA}
The set of possible values for Alice's dataset, given Eve's observations, is 
\[ \Pos(X_A) \triangleq \{\Bar{X} \in \mathcal{X}^{|X_A|} :  \exists T \in \mathcal{F}_{T_A} \text{ with } T(\Bar{X}) = T_A(X_A)\} .\]
\end{proposition}

The next theorem characterizes what Eve learns about $\bm{X_A}$ via Alice's scheme and what she would learn if, instead, she only observed $Y_A = \{L(x)\}_{x\in\ X_A}$.

\begin{theorem} \label{theorem: probabilities}
Let $\bm{X_A}$ be the random variable for Alice's private samples and $\bm{T_A}$ be the random variable for Alice's encoder. We note $\mathbbm{1}(\text{condition})$ is an indicator function which takes value one when condition happens and zero otherwise. Then,
\begin{equation*}
    \begin{split}
        \Pr[\bm{X_A} &= X_A \mid \bm{T_A ({X}_A)}=Z,\; \bm{\LC(T_A)}=C]\propto \hspace{-0.3cm}\sum_{\{T:\LC(T)=C,T(X_A)=Z\}}\hspace{-0.6cm}\Pr[\bm{{X}_A}=X_A]\Pr{[ \bm{T_A} = T]}
    \end{split}
\end{equation*}
and
\[\Pr[\bm{X_A} = X_A \mid \bm{Y_A}=Y_A] \propto \mathbbm{1} (Y_A = \{L(x)\}_{x\in\ X_A} ) \cdot \Pr[\bm{X_A} = X_A] .\]
\end{theorem}

Since Theorem~\ref{theorem: probabilities} fully characterizes how much Eve learns about Alice's private samples and the encoder $T_A$, any privacy metric, be it a mutual information \citep{Shannon48}, a mean squared error, or a guessing measure \citep{pliam1999guesswork}, can be computed from it.

The following example showcases the theorem.

\begin{example}
Let $\mathcal{X}=\{1,2,3,4,5\}$, $X_A = \{2,3,4\}$, $\LC(\mathcal{X}) =(++---)$, and $\mathcal{F}=\{T_1,\dots,T_6\}$ given below. Suppose $T_A = T_1$. Then, Alice transmits $\{(1,+),(4,-),(5,-)\}$. Eve knows $\LC(T_A) =(++---)$, and thus knows $\mathcal{F}_{T_A}$ as well:
\begin{equation*}
\mathcal{F}=\begin{cases}
T_1:&(2,1,5,4,3)\\ 
T_2:&(2,1,3,5,4)\\ 
T_3:&(1,2,3,5,4)\\ 
T_4:&(4,3,1,2,5)\\ 
T_5:&(3,4,2,1,5)\\ 
T_6:&(5,2,4,3,1)
\end{cases},\quad
\mathcal{F}_{T_A}=\begin{cases}
T_1:&(2,1,5,4,3)\\ 
T_2:&(2,1,3,5,4)\\ 
T_3:&(1,2,3,5,4)\\ 
\end{cases}.
\end{equation*}
As a result, $\Pos (X_A) = \{ (2,4,3), (2,5,4), (1,5,4) \}$. Thus, Eve knows that $4 \in X_A$ independent of the distribution of $X_A$. If we assume uniform distributions for $T_A$ and $X_A$, then, by Theorem \ref{theorem: probabilities}, we obtain a uniform distribution on $\Pos (X_A)$ and can compute $\Pr[1 \in X_A]={1}/{3}$ and $\Pr[2 \in X_A] = {2}/{3}$. If instead we assume uniform $T_A$ and $\Pr[X_A=\{2,3,4\}] = 0.1$, $\Pr[X_A=\{2,5,4\}] = 0.1$, and $\Pr[X_A=\{1,5,4\}] = 0.4$, then $\Pr[1 \in X_A] = {2}/{3}$ and $\Pr[2 \in X_A] = {1}/{3}$. A more detailed version of this example is given in the appendix.
\label{exmaple2:extensive}
\end{example}

We now describe a scheme which achieves perfect privacy without affecting learnability. 

\begin{theorem} \label{theorem: optimal scheme}
Let $\mathcal{X}^y = \{x \in \mathcal{X}: L(x) = y \}$. Then, sampling $T_A$ uniformly from the family $\mathcal{F}_0 = \{T \in \Sym (\mathcal{X}) : T(\mathcal{X}^y) = \mathcal{X}^y \quad \forall y \in \mathcal{Y} \}$ achieves perfect privacy without altering the structure of the labels.
\end{theorem}

Although Theorem \ref{theorem: optimal scheme} shows the existence of an optimal family of encoders, Alice has no way of sampling from it as it would require full knowledge of $\mathcal{X}$ and its labels. Thus, the theorem serves as a guide on what properties Alice might want from the family of encoders $\mathcal{F}$. 

Starting from an imperfect $\mathcal{F}$, we now wish to understand what kind of operations Alice can perform to enrich the privacy of $\mathcal{F}$. To this end, we explore several ways to grow $\mathcal{F}$. In our next proposition, we show that adding arbitrary functions to $\mathcal{F}$ might actually worsen the privacy.


\begin{proposition} \label{prop:addingcanhurt} Let $\mathcal{F}, \mathcal{F}' \subseteq \Sym (\mathcal{X})$ be two families of encoders such that $\mathcal{F} \subseteq \mathcal{F}'$. Then, it is not generally true that $\mathcal{F}'$ is more private than $\mathcal{F}$.
\end{proposition}

However, as we now show, composing families of functions can only preserve or increase the privacy of the family.

\begin{theorem} \label{theorem:compositiondoesnothusrt}
Let $\mathcal{F}, \mathcal{F}' \subseteq \Sym (\mathcal{X})$ and $\mathcal{F}' \circ \mathcal{F} = \{ T' \circ T : T' \in \mathcal{F}', T \in \mathcal{F} \}$. Then, $\mathcal{F}' \circ \mathcal{F}$ is no less private than $\mathcal{F}$.
\end{theorem}

Theorem \ref{theorem:compositiondoesnothusrt} shows that composing families of encoders cannot reduce the privacy. Indeed, as we show in the following example, it can potentially increase it.

\begin{example}
\label{ex:compositionhelp}
Let $\mathcal{X}=\{1,2,3,4,5\}$ and $\LC(\mathcal{X})=(++---)$. The sets of encoders $\mathcal{F}$, $\mathcal{F}'$, and  $\mathcal{F}\circ\mathcal{F}'$, along with the label configuration for each encoder, are given:

\begin{minipage}[c]{0.45\linewidth}
     \begin{equation*}
        \mathcal{F}=\begin{cases}
        T_1:&(1,2,3,4,5),(++---)\\ 
        T_2:&(2,1,3,5,4),(++---)\\ 
        T_3:&(1,2,5,4,3),(++---)\\ 
        T_4:&(1,3,2,4,5),(+-+--)\\ 
        T_5:&(1,5,2,3,4),(+-+--)
        \end{cases}
    \end{equation*}
        \begin{equation*}
        \mathcal{F'}=\begin{cases}
        T'_1:&(1,2,3,4,5),(++---)\\ 
        T_2:&(3,2,1,4,5),(-++--) 
        \end{cases}
    \end{equation*}
\end{minipage}
\begin{minipage}[c]{0.52\linewidth}
    \begin{equation*}
    \mathcal{F'}\circ\mathcal{F}=\begin{cases}
        T'_1\circ T_1:&(1,2,3,4,5),(++---)\\ 
        T'_1\circ T_2:&(2,1,3,5,4),(++---)\\ 
        T'_1\circ T_3:&(1,2,5,4,3),(++---)\\ 
        T'_1\circ T_4:&(1,3,2,4,5),(+-+--)\\ 
        T'_1\circ T_5:&(1,5,2,3,4),(+-+--)\\ 
        T'_2\circ T_1:&(3,2,1,4,5),(-++--)\\ 
        T'_2\circ T_2:&(3,1,2,5,4),(-++--)\\ 
        T'_2\circ    T_3:&(5,2,1,4,3),(-++--)\\ 
        T'_2\circ T_4:&(2,3,1,4,5),(+-+--)\\ 
        T'_2\circ          T_5:&(2,5,1,3,4),(+-+--) 
        \end{cases}
    \end{equation*}
\end{minipage}

Here, $\mathcal{F}$ has two LC-anonymity lists, with cardinality $2$ and $3$, and $\mathcal{F}'\circ\mathcal{F}$ has three LC-anonymity lists with cardinality $3$, $3$, and $4$. Thus, it offers more ambiguity (better privacy) as its LC-anonymity lists have higher minimum cardinality compared to $\mathcal{F}$ alone.
\end{example}

While we cannot directly sample from the optimal family of encoders, we can leverage our theoretical results on function composition to guide the design of \name.  Starting with a weak encoder, i.e a linear layer, we iteratively enrich the privacy of our function family through function composition (e.g  by adding with additional non-linear and linear layers), to build a random neural network. Moreover, given that we know that the labeling function $L_A$ on medical images is likely to be efficiently estimated with a convolutional neural network, we implement our \name encoders as convolutional neural networks. We emphasize that our privacy guarantees for the optimal family do not extend to our implementation of \name, and thus its privacy must be tested empirically, following standard practice in cryptanalysis \citep{standard2001announcing, dworkin2015sha}.

%% file: experiments.tex
\textbf{Datasets.} For all experiments, we utilized two benchmark datasets of chest x-rays, MIMIC-CXR (\cite{johnson2019mimic}) and  CheXpert (\cite{irvin2019chexpert}) from Beth Israel Deaconess Medical Center and Stanford respectively. For each dataset, we evaluated the ability of each model to predict Edema, Pneumothorax, Consolidation, Cardiomegaly and Atelectasis. For each task, we excluded exams with an uncertain disease label, i.e., the clinical diagnosis did not explicitly rule out or confirm the disease, and randomly split the remaining data $60{-}20{-}20$ for training, development and testing respectively. All images were down sampled to $256{\times}256$ pixels. All experiments were repeated 3 times across different seeds and we report each metric with its standard deviation.

\subsection{Evaluating modeling utility}
To evaluate the impact of \name encodings on downstream modeling performance, we compared \name-based models to standard architectures trained on raw images across both the single and multi-hospital setting. For each diagnosis task and training setting, we report the average AUC across the MIMIC-CXR and CheXpert test sets. We report results on the MIMIC-CXR and CheXpert datasets individually in the appendix. For \name multi-hospital training, we wished to evaluate the impact of leveraging independent \name encoders on modeling accuracy. As a result, we evaluate both model performance when leveraging a single encoder across both hospitals (Combined-Clear), and model performance when leveraging two independent encoders (Combined-Private). We note that performance in the Combined-Clear setting acts as an upper-bound for Combined-Private.  

Our \name encoding leveraged a patch-size of $16{\times}16$, a depth of $7$, and a hidden dimension of $2048$. This model had $\sim22.9M$ parameters and mapped $256{\times}256$ pixel images to $256{\times}2048$ vectors. Due to the patch-shuffling step in \name, this representation is unordered. As a result, we trained Vision Transformers (ViT) \citep{zhou2021deepvit}, a self-attention based architecture that is invariant to patch ordering. Across all experiments, we used a one-layer ViT with a hidden dimension of $2048$.  We compared \name model performance with a non-private baseline, namely an identical ViT model where the patch encoder is learned jointly. We trained all models for $25$ epochs using the Adam optimizer \citep{kingma2014adam}, an initial learning rate of $1\mathrm{e}{-04}$, weight decay of $1\mathrm{e}{-03}$ and a batch size of $128$.

\subsection{Evaluating robustness to attacks}

\textbf{Adversarial attack.} To validate robustness of our encoding approach to attacks aimed at estimating $T$, we preformed experiments on the combined MIMIC-CXR and CheXpert datasets with Cardiomegaly labels. We assumed that the attacker has access to the entire labeled dataset $\{(x_i,y_i)\}_{x_i\in X}$, and labeled \name-encoded samples $\{(z_i=T(x_i),y_i)\}_{x_i\in X}$. We also assumed that the attacker knows $\mathcal{F}$ (the exact architecture of \name), but not the weights of $T$ which are private. Given this information, the attacker tries to learn a $T^*$ such that $T(X) \approx T^*(X)$. We assume that if $T$ was leaked, the attacker could easily invert the encoding and recover the raw original images. To estimate $T$, we sampled an initial  $T^*$ with the same architecture as $T$, and trained it to minimize the accuracy of a domain discriminator, which aims to distinguish between the generated $Z^*$ and true $Z$. 

The discriminator is typically designed as a parameterized classifier \citep{tzeng2017adversarial, goodfellow2014generative, shen2017style}, and trained through a mini-max game with the encoder $T^*$. However, we found this difficult to train. Instead, we conducted experiments using Maximum Mean Discrepancy (MMD) \citep{gretton2012kernel} as our discriminator. Specifically, we defined the MMD loss as for a batch of real ciphertext and generated ciphertext $Z$ and $Z^*$ as $ L_{T^*} = \text{MMD}^2(Z, Z^*)$, where
$\text{MMD}(Z, Z^*) = \left\| (\sum_z \phi(z) )/|Z|-  (\sum_{z^*} \phi(z^*))/| Z^*|   \right\|_\mathcal{H}$
measures the discrepancy between $Z$ and $Z^*$ on a Reproducing Kernel Hilbert Space, and $\phi$ is a feature map induced by a linear combination of multiple RBF kernels $\kappa(z_i, z_j) = \sum_n \exp( -{1}/(2\sigma_n \left \| z_i - z_j \right \|^2))$. Our MMD formulation follows prior work in domain adaptation \citep{guo2018multi, bousmalis2016domain}. 

In order to understand how the success of this attack would vary with the architecture of \name, we performed the attack on \name architectures with a depth of $2$ and $7$. As a baseline, we also performed the attack when using a simple linear encoder, implemented as single convolutional layer. Across all experiments, we used a hidden dimension of $2048$ and trained $T^*$ for $25$ epochs. We performed a grid search over different learning rates and weight decay values for each attack. For each experiment, we evaluated the attack by measuring the mean squared error (MSE) between generated ($Z^*$) and real ciphertext ($Z$) for the same plaintext images across the dataset. To understand if $T^*$ outperforms a trivial baseline, we compare the performance of $T^*$ to the mean baseline $T_\mu$, where $T_\mu(x) = \frac{1}{|Z|} (\sum_z z)$. $T_\mu$ ignores the input $x$ and predicts the mean of $Z$ for all inputs. We consider an attack to be successful if $T^*$ outperforms $T_\mu$, i.e., the ratio of $\frac{T^* \text{MSE}}{T^\mu \text{MSE}} < 1$. We consider additional experiments varying the architecture of $T$ and $T^*$ and provide additional analyses in the appendix. 

\textbf{Transfer learning attack.} We also consider a scenario where an attacker may try to learn a sensitive attribute classifier, such as a gender predictor, on chest x-rays and tries to transfer this classifier onto the encoded data $Z$. If this transferred classifier performs better than random on $Z$, then an attacker, could leverage this approach to learn sensitive information not released by the data-owner or to propagate labels to be used in a refined adversarial attack.  To validate the robustness of \name to this type of attack, we began with the best estimated $T^*$ from our adversarial attack experiments, as measured by $T^*$ MSE, and built a new classifier to predict Edema given $Z^*$. We report the ROC AUC of this classifier on both $Z$ and $Z^*$, We performed these experiments on the combined MIMIC-CXR and CheXpert dataset.  We performed this attack for both \name-encoding with a depth of $7$ and a hidden dimension of $2048$, as leveraged in the \textit{modeling utility} experiments, and when using a linear encoding. For each experiment, we trained a ViT for 25 epochs using the Adam optimizer, an initial learning rate of $1\mathrm{e}{-04}$ and a batch size of $128$.

%% file: results.tex
\input{Tables/diagnosis}

\textbf{Evaluating modeling utility.}
We report our results in predicting various medical diagnoses from chest x-ray datasets in Table \ref{tab:xray-expers}. \name-ViT obtained competitive AUCs to our non-private ViT baseline across all training settings. In the multi-hospital setting, we found that \name-ViT was effectively able to leverage the larger training set to learn an improved classifier, despite using separate encoders for each dataset. \name-ViT obtained an average AUC increase of 2 and 3 points compared to training only on the MIMIC-CXR and CheXpert datasets respectively. Moreover, \name demonstrated achieved equivalent performance in the Combined-Clear and Combined-Private settings, demonstrating that multiple institutions do not pay a significant performance cost to collaborate privately.

\input{Tables/attack}

\textbf{Evaluating robustness to attacks.} We report the performance of our adversarial and transfer learning attacks in Table~\ref{tab:attack} left and right panels 
respectively. As expected, using a linear encoding is not robust to either adversarial or transfer learning based attacks. We found that the adversarial attack outperformed the mean baseline (i.e. $\frac{T^* \text{MSE}}{T^\mu \text{MSE}} < 1$), and an Edema classifier trained on $T^*$ transferred well to $T$ encodings. In contrast, our adversarial attack on \name failed to outperform the mean baseline when both using $2$ and $7$ layers. Moreover, \name was robust to our transfer learning attack, with the Edema classifier failing to obtain a AUC significantly better than random on the true $T$ encodings.

%% file: Tables/diagnosis.tex
\begin{table}[!htb]
\centering
\caption{Impact of \name on chest x-ray prediction tasks across different training settings. All metrics are average ROC AUCs across the MIMIC-CXR and CheXpert test sets. Combined-Clear and Combined-Private refer to using a single \name encoder across the combined MIMIC-CXR and CheXpert datasets and two independent \name encoders respectively. Guides of abbreviations for medical diagnosis: (E)dema, (P)neumothorax, (Co)nsolidation, (Ca)rdiomegaly and (A)telectasis.}
\label{tab:xray-expers}
\begin{tabular}{ccccccc}
\hline
\textit{Model} & E & P & Co & Ca & A & \textit{Average}\\ 
\hline
\multicolumn{7}{c}{\textbf{Train on MIMIC-CXR}}  \\
\hline
ViT & 85 $\pm $ 1  & 69 $\pm $ 3 & 74$ \pm $ 2 & 87 $\pm $ 0 & 83 $\pm $ 1 & \textit{80} \\
\hline
\name-ViT & 85 $\pm $ 2 & 72 $\pm $ 1 & 72 $\pm $ 1 & 87 $\pm $ 0 & 83 $\pm $ 1 & \textit{80} \\
\hline
\multicolumn{7}{c}{\textbf{Train on CheXpert}} \\
\hline
ViT & 82 $\pm $ 1 & 71 $\pm $ 1 & 72 $\pm $ 3 & 83 $\pm $ 1 & 80 $\pm $ 0 & \textit{77} \\
\hline
\name-ViT & 84 $\pm $ 1 & 71 $\pm $ 1 & 75 $\pm $ 2 & 82 $\pm $ 1 & 81 $\pm $ 0 & \textit{79} \\
\hline
\multicolumn{7}{c}{\textbf{Train on Combined-Clear}} \\
\hline
ViT & 86 $\pm $ 0 & 77 $\pm $ 1 & 76 $\pm $ 2 & 87 $\pm $ 1 & 85 $\pm $ 0 & \textit{82} \\
\hline
\name-ViT & 87 $\pm $ 0 & 76 $\pm $ 3 & 78 $\pm $ 1 & 88 $\pm $ 0 & 85 $\pm $ 1 & \textit{83} \\
\hline
\multicolumn{7}{c}{\textbf{Train on Combined-Private}} \\
\hline
\name-ViT & 87 $\pm $ 1 & 77$\pm $ 3 & 77 $\pm $ 3 & 86 $\pm $ 1 & 84 $\pm $ 1 & \textit{82}\\
\hline
\end{tabular}
\end{table}


%% file: Tables/attack.tex
\begin{table}[!htb]
\centering
\caption{Left: MSE of MMD-based adversarial attacks on different \name encodings. $T_\mu$ refers to the mean baseline, and $T^*$ refers to the encoder learned via the MMD attack. Right: Performance of transfer learning attack on Linear and \name encodings.}
\label{tab:attack}
\begin{tabular}{ c c}
\hline
Encoding &  $T^* \text{MSE} / T_{\mu} \text{MSE} $ \\ 
\hline
Linear &  0.43 $\pm$ 0.01  \\ 
\hline
\name-depth-2  & 7.97 $\pm$ 0.28\\ 
\hline
\name-depth-7 & 4.44 $\pm$ 0.12    \\
\hline
\end{tabular}
\hspace{0.1cm}
\begin{tabular}{ccc}
\hline
Encoding & $T^*$ AUC & $T$ AUC\\ 
\hline
Linear & 89 $\pm$ 1 & 86 $\pm$ 1 \\ 
\hline
\name & 84 $\pm$ 1 & 52 $\pm$ 4 \\ 
\hline
\end{tabular}

\end{table}


%% file: discussion.tex
We proposed \name, a private encoding scheme based on random neural networks designed to enable data owners to  publicly publish their datasets while retaining data privacy and modeling utility. On two benchmark chest x-ray datasets, MIMIC-CXR and CheXpert, we found that \name-models obtained competitive performance to our non-private baselines. In the multi-institutional setting, where each site leverages an independent name encoder, we demonstrated that \name-models could effectively leverage the larger training data to learn improved classifiers. While this paper focused on medical imaging and chest X-ray tasks, \name can easily be extended to new data modalities such as text or molecular graphs.  While we are not able to sample from the optimal family of encoders identified in our theoretical analysis, our analysis also provided a useful guide (i.e. function composition) for the design of \name. Similar to prior work in cryptanalysis \citep{standard2001announcing, dworkin2015sha}, we note that our empirical results on adversarial robustness are not sufficient to prove the privacy of our architecture family for \name. Improved algorithms for domain adaption \citep{tzeng2017adversarial,guo2018multi} or unsupervised translation \citep{lample2017unsupervised, alvarez2018gromov} specialized to the design of \name may yield more successful attacks. We release a challenge dataset to both encourage the development of new attacks on \name as well as the development of improved \name architectures.

%% file: impact.tex
Our study proposes a method to enable data-owners, e.g hospitals, to share their data publicly while protecting both patient privacy and modeling utility. We hope that this technology will enable the construction of diverse multi-center patient cohorts, and allow the broader machine learning community to contribute to the development of healthcare algorithms. We believe that improved algorithms in this space will lead to more equitable and precise healthcare. However, we acknowledge that the same privacy-preserving technology could be used to accelerate unethical biomedical research. 

%% file: appendixA.tex
In this section, we prove the results of Section~4 and give an extended version of Example 2. We begin by proving the following lemma.

\begin{lemma} \label{lemma:probAlice}
Let $\bm{X_A}$ be the random variable for Alice's private samples and $\bm{T_A}$ be the random variable for Alice's encoder. Then, $\Pr[\bm{X_A}= X_A , \bm{T_A} = T \mid \bm{T _A({X_A})}=Z,\; \bm{\LC(T_A)}=C]$ is proportional to $\mathbbm{1} (T(X_A)=Z) \cdot \mathbbm{1}(\LC(T)=C) \cdot \Pr[\bm{{X}_A}=X_A] \cdot \Pr{[ \bm{T_A} = T]}$.
\end{lemma}

\paragraph{Proof of Lemma~\ref{lemma:probAlice}.}
Let $P_{1} = \Pr[\bm{X_A} = X_A,\bm{T_A} = T \mid \bm{T_A({X}_A)}=Z,\; \bm{\LC(T_A)}=C]$ be the conditional joint distribution of $\bm{X_A}$ and $\bm{T_A}$, $P_{2} = \Pr[\bm{X_A} = X_A \mid \bm{T_A} = T, \bm{T_A ({X}_A)}=Z,\; \bm{\LC(T_A)}=C]$, and $P_3= \Pr[\bm{T_A} = T\mid \bm{T_A ({X}_A)}=Z,\; \bm{\LC(T_A)}=C]$. Then, $P_1 = P_2 P_3$, and a direct calculation shows that $P_2 =\mathbbm{1} (T(X_A)=Z)$. As for $P_3$, it follows from Bayes' theorem that
\begin{align} \label{eq: p3}
    P_3 \propto  \Pr[\bm{T_A ({X}_A)}=Z\mid\;\bm{T_A} = T,\; \bm{\LC(T_A)}=C] \cdot \Pr[\bm{T_A} = T,\; \bm{\LC(T_A)}=C].
\end{align}
Next, we have $\Pr[\bm{T_A ({X}_A)}=Z\mid\;\bm{T_A} = T,\; \bm{\LC(T_A)}=C] = \Pr[\bm{{X}_A}=T^{-1}(Z)]$. Also, by Bayes' theorem, $\Pr[\bm{T_A} = T,\; \bm{\LC(T_A)}=C] \propto \Pr[ \bm{\LC(T_A)}=C|\bm{T_A} = T] \cdot \Pr[ \bm{T_A} = T]$. Finally, from a direct calculation we obtain $\Pr[ \bm{\LC(T_A)}=C|\bm{T_A} = T] = \mathbbm{1} (\LC(T)=C)$. The result follows from substituting everything into \eqref{eq: p3} and multiplying by $P_2$. We note that $\mathbbm{1} (T(X_A)=Z) \cdot \Pr[\bm{{X}_A}=T^{-1}(Z)] = \mathbbm{1} (T(X_A)=Z) \cdot \Pr[\bm{{X}_A}=X_A]$.
\qed

\paragraph{Theorem~1.}
Let $\bm{X_A}$ be the random variable for Alice's private samples and $\bm{T_A}$ be the random variable for Alice's encoder. Then,
\begin{equation*}
    \begin{split}
        \Pr[\bm{X_A} &= X_A \mid \bm{T_A ({X}_A)}=Z,\; \bm{\LC(T_A)}=C]\propto \hspace{-0.3cm}\sum_{\{T:\LC(T)=C,T(X_A)=Z\}}\hspace{-0.6cm}\Pr[\bm{{X}_A}=X_A]\Pr{[ \bm{T_A} = T]}
    \end{split}
\end{equation*}
and
\[\Pr[\bm{X_A} = X_A \mid \bm{Y_A}=Y_A] \propto \mathbbm{1} (Y_A = \{L(x)\}_{x\in\ X_A} ) \cdot \Pr[\bm{X_A} = X_A] .\]

\paragraph{Proof of Theorem~1.}
For the first statement,
\begin{align}
    P_A &= \sum_{T}\Pr[\bm{X_A}= X_A,\bm{T_A} = T \mid \bm{T_A({X}_A)}=Z,\; \bm{\LC(T_A})=C] \label{eq:teo1 1}\\
    &\propto \sum_{T}\mathbbm{1} (T(X_A)=Z)\cdot\mathbbm{1} (\LC(T)=C)\cdot\Pr[\bm{{X}_A}=X_A]\cdot\Pr{[ \bm{T_A} = T]} \label{eq:teo1 2}\\
    &= \sum_{\{T:\LC(T)=C,T(X_A)=Z\}}\hspace{-0.6cm}\Pr[\bm{{X}_A}=X_A]\cdot\Pr{[ \bm{T_A} = T]}, \label{eq:teo1 3}
\end{align}
where \eqref{eq:teo1 1} follows from obtaining the marginal distribution from the joint distribution, and \eqref{eq:teo1 2} follows from Lemma \ref{lemma:probAlice}.

As for the second statement, from an application of Bayes' theorem, we obtain
\begin{align*}
    \Pr[\bm{X_A} = X_A \mid \bm{Y_A}=Y_A] &\propto \Pr[\bm{Y_A} = Y_A \mid \bm{X_A}=X_A].\Pr[\bm{X_A}=X_A]\\
        &\propto\mathbbm{1} (Y_A = \{L(x)\}_{x\in\ X_A} ). \Pr[\bm{X_A} = X_A].
\end{align*}
\qed

\paragraph{Proposition~1.}
The set of possible values for Alice's dataset, given Eve's observations, is 
\[ \Pos(X_A) \triangleq \{\Bar{X} :  \exists T \in \mathcal{F}_{T_A} \text{ with } T(\Bar{X}) = T_A(X_A)\} .\]

\paragraph{Proof of Proposition~1.} Let $Z=T_A(X_A)$, $C=\LC(T_A)$, and $X \notin \Pos(X_A)$. Then, there does not exist a $T \in \mathcal{F}_A$ such that $T(X)=Z$. But $T \in \mathcal{F}_A$ if and only if $\LC(T)=C$. Thus, $\{T:\LC(T)=C,T(X)=Z\}=\varnothing$, and therefore, it follows from Theorem~1 that $$\Pr[\bm{X_A} = X\mid \bm{T_A({X}_A)}=Z,\; \bm{\LC(T_A})=C] = 0.$$
\qed

\paragraph{Extended version of Example~2.}
Let $\mathcal{X}=\{1,2,3,4,5\}$, $X_A = \{2,3,4\}$, $\LC(\mathcal{X}) =(++---)$, and
\begin{equation*}
\mathcal{F}=\begin{cases}
T_1:&(2,1,5,4,3)\\ 
T_2:&(2,1,3,5,4)\\ 
T_3:&(1,2,3,5,4)\\ 
T_4:&(4,3,1,2,5)\\ 
T_5:&(3,4,2,1,5)\\ 
T_6:&(5,2,4,3,1)
\end{cases}.
\end{equation*}

\begin{itemize}
    \item Suppose $T_A = T_1$. Then, Alice transmits $\{(1,+),(5,-),(4,-)\}$. Under our threat model, we assume that Eve knows $\LC(T_A) =(++---)$, and therefore, that $T_A$ is in the set
\begin{equation*}
\mathcal{F}_{T_A}=\begin{cases}
T_1:&(2,1,5,4,3)\\ 
T_2:&(2,1,3,5,4)\\ 
T_3:&(1,2,3,5,4)\\ 
\end{cases} .
\end{equation*}
which implies that $X_A \in \Pos (X_A) = \{ (2,3,4), (2,4,5), (1,4,5) \}$. Thus, Eve knows that $4 \in X_A$ independent of any distribution for $X_A$. As for the the two other values, it follows from Theorem~1 that
{
\scriptsize
\begin{equation*}
    \begin{split}
        &\Pr[1 \in \bm{X_A}]=\sum_{\{\Bar{X}:1\in\Bar{X}\}}\Pr[\bm{X_A}=\Bar{X}\mid \bm{T_A({X}_A)}=(1,5,4),\; \bm{\LC(T_A})=(++---)]\\
        &=\frac{\Pr[\bm{{X}_A}=(1,4,5)]\Pr{[ \bm{T_A} = T_3]}}{\Pr[\bm{{X}_A}=(2,3,4)]\Pr{[ \bm{T_A} = T_1]}+\Pr[\bm{{X}_A}=(2,4,5)]\Pr{[ \bm{T_A} = T_2]}+\Pr[\bm{{X}_A}=(1,4,5)]\Pr{[ \bm{T_A} = T_3]}}
    \end{split}
\end{equation*}
}
and
{
\scriptsize
\begin{equation*}
    \begin{split}
        &\Pr[2 \in \bm{X_A}]=\sum_{\{\Bar{X}:2\in\Bar{X}\}}\Pr[\bm{X_A}=\Bar{X}\mid \bm{T_A({X}_A)}=(1,5,4),\; \bm{\LC(T_A})=(++---)]\\
        &=\frac{\Pr[\bm{{X}_A}=(2,3,4)]\Pr{[ \bm{T_A} = T_1]}+\Pr[\bm{{X}_A}=(2,4,5)]\Pr{[ \bm{T_A} = T_2]}}{\Pr[\bm{{X}_A}=(2,3,4)]\Pr{[ \bm{T_A} = T_1]}+\Pr[\bm{{X}_A}=(2,4,5)]\Pr{[ \bm{T_A} = T_2]}+\Pr[\bm{{X}_A}=(1,4,5)]\Pr{[ \bm{T_A} = T_3]}}
    \end{split}
\end{equation*}
}
If we assume uniform distributions for $T_A$ and uniform and iid distribution for samples in $X_A$, then $\Pr[1 \in X_A]={1}/{3}$ and $\Pr[2 \in X_A] = {2}/{3}$. If we instead assume uniform $T_A$ and that $\Pr[X_A=\{2,3,4\}] = 0.1$, $\Pr[X_A=\{2,4,5\}] = 0.1$, and $\Pr[X_A=\{1,4,5\}] = 0.4$, then $\Pr[1 \in X_A] = {2}/{3}$ and $\Pr[2 \in X_A] = {1}/{3}$.

\item Suppose $T_A = T_4$. Then, Alice transmits $\{(3,+),(1,-),(2,-)\}$. Eve knows $\LC(T_A)=(--++-)$, and 
\begin{equation*}
\mathcal{F}_{T_A}=\begin{cases}
T_4:&(4,3,1,2,5)\\ 
T_5:&(3,4,2,1,5)
\end{cases}
\end{equation*}
and $\Pos (X_A) = \{ (2,3,4), (1,4,3) \}$. Thus, Eve knows that $\{3,4\}\subset X_A$. The other probabilities can be derived in a similar way to the above item.
\item Suppose $T_A = T_6$. Then, Alice transmits $\{(2,+),(4,-),(3,-)\}$. Eve knows $\LC(T_A)=(-+--+)$, and \begin{equation*}
\mathcal{F}_{T_A}=\begin{cases}
T_6:&(5,2,4,3,1)
\end{cases}
\end{equation*} 
and, therefore, $\Pos (X_A) = \{ (2,3,4) \}$. In this case, Eve can exactly determine $X_A$.
\end{itemize}

\paragraph{Theorem~2.}
Let $\mathcal{X}^y = \{x \in \mathcal{X}: L(x) = y \}$. Then, sampling $T_A$ uniformly from the family $\mathcal{F}_0 = \{T \in \Sym (\mathcal{X}) : T(\mathcal{X}^y) = \mathcal{X}^y \quad \forall y \in \mathcal{Y} \}$ achieves perfect privacy without altering the structure of the labels.

\paragraph{Proof of Theorem~2.} 
Let $Z=T_A(X_A)$ and $C=\LC(T_A)$. We note that every $T \in \mathcal{F}_0$ has the same label configuration, i.e., $\LC (T) = \LC (\mathcal{X})=C$. It follows then from Theorem~1 that
\begin{multline*} 
    \Pr[\bm{X_A} = X \mid \bm{T_A({X}_A)}=Z,\; \bm{\LC(T_A})=C] 
    \\ = \sum_{T\in \mathcal{F}_0} \mathbbm{1} (T(X)=Z) \cdot \mathbbm{1}(\LC(\mathcal{X})=C) \cdot \Pr[\bm{{X}_A}=X] \cdot \Pr{[ \bm{T_A} = T]}  \\
    = \sum_{T\in \mathcal{F}_0} \mathbbm{1} (T(X)=Z)\cdot \Pr[\bm{{X}_A}=X] \cdot \Pr{[ \bm{T_A} = T]} .
\end{multline*}
Since $T$ is chosen uniformly from $\mathcal{F}_0$, it follows that
\begin{multline} \label{eq: teo2 1}
\Pr[\bm{X_A} = X \mid \bm{T_A({X}_A)}=Z,\; \bm{\LC(T_A})=C]\\
\propto\Pr[\bm{{X}_A}=X]\sum_{T\in \mathcal{F}_0} \mathbbm{1}(T(X)=Z) = \Pr[\bm{{X}_A}=X]\cdot{|\{ T \in \mathcal{F}_0 : T(X)=Z \}|} .
\end{multline}

Since Eve knows the set $Z$ and the label configuration $C$, then she also knows the set of labels $Y_A = \{L(x)\}_{x\in\ X_A}$. The reason for this is that $C$ is the vector representation of the labeling function $L\circ T^{-1}$, and thus, since Eve knows $C$ and $Z$, she also knows $Y_A=L\circ T^{-1}(Z)$. We also note that the equality $\{L(x)\}_{x\in\ X}=Y_A$ holds if and only if the labels of $X$ and $Z$ are in agreement, i.e., $|\{x\in X:L(x)=y\}|=|\{z\in Z:L\circ T^{-1}(z)=y\}|$ for every $y\in\mathcal{Y}$. This is equivalent to the relation
\[ \mathbbm{1}(Y_A=\{L(x)\}_{x\in X}) = \prod_{y \in \mathcal{Y}} \mathbbm{1}(|X^y|=|Z^y|), \]
where $X^y = \{ x \in X : L(x)=y \}$ and $Z^y = \{z\in Z:L\circ T^{-1}(z)=y\}$.

On the other hand, 
\[ |\{ T \in \mathcal{F}_0 : T(X)=Z \}| = \prod_{y \in \mathcal{Y}} |X^y|! \cdot  \mathbbm{1}(|X^y|=|Z^y|).\]
We note that $|X^{y}|$ does not depend on $X$ and is in fact given by the number of elements with value $y$ in $Y_A$. Thus, 
\[ |\{ T \in \mathcal{F}_0 : T(X)=Z \}| \propto \prod_{y \in \mathcal{Y}}  \mathbbm{1}(|X^y|=|Z^y|) = \mathbbm{1}(Y_A=\{L(x)\}_{x\in X}).\]

Therefore, by substituting in \eqref{eq: teo2 1}, we obtain
\[\Pr[\bm{X_A} = X \mid \bm{T_A({X}_A)}=Z,\; \bm{\LC(T_A})=C] \propto \mathbbm{1}(Y_A=\{L(x)\}_{x\in X})  \Pr[\bm{X_A}=X].\]

Then, by Theorem~1, 
\[ \Pr[\bm{X_A} = X \mid \bm{T_A({X}_A)}=Z,\; \bm{\LC(T_A})=C] = \Pr[\bm{X_A} = X \mid \bm{Y_A}=Y_A] .\]
In other words, sampling $T$ from $\mathcal{F}_0$ uniformly at random is perfectly private. The fact that it does not alter the structure of the labels follows directly from $\LC (T) = \LC (\mathcal{X})$ for every $T \in \mathcal{F}_0$.
\qed

\paragraph{Proposition~ 2.}
Let $\mathcal{F}, \mathcal{F}' \subseteq \Sym (\mathcal{X})$ be two families of encoders such that $\mathcal{F} \subseteq \mathcal{F}'$. Then, it is not generally true that $\mathcal{F}'$ is more private than $\mathcal{F}$.

\paragraph{Proof of Proposition~2.}
Let $\mathcal{X}=\{1,2,3,4\}$ with $\LC(\mathcal{X})=(++--)$,
\begin{equation*}
\mathcal{F}=\begin{cases}
T_1:&(1,2,3,4)\\ 
T_2:&(2,1,3,4)\\
T_3:&(1,2,4,3)\\ 
T_4:&(2,1,4,3)
\end{cases},\quad \text{and} \quad
\mathcal{F}'=\begin{cases}
T_1:&(1,2,3,4)\\ 
T_2:&(2,1,3,4)\\
T_3:&(1,2,4,3)\\ 
T_4:&(2,1,4,3)\\ 
T_5:&(3,4,1,2)
\end{cases}
\end{equation*}
Then, $\mathcal{F}$ is \textit{optimal}; indeed it is equal to $\mathcal{F}_0$ as defined in Theorem~2. But $\mathcal{F}'$ is not, as it has an LC-anonimity list with a single member, $T_5$. In fact, whenever $T_5$ is selected from $\mathcal{F}'$, Eve can perfectly decode $X_A$.
\qed

\paragraph{Theorem~3.}
Let $\mathcal{F}, \mathcal{F}' \subseteq \Sym (\mathcal{X})$ and $\mathcal{F}' \circ \mathcal{F} = \{ T' \circ T : T' \in \mathcal{F}', T \in \mathcal{F} \}$. Then, $\mathcal{F}' \circ \mathcal{F}$ is no less private than $\mathcal{F}$.

\paragraph{Proof of Theorem~3.}

Let $T'_A \circ T_A \in \mathcal{F}' \circ \mathcal{F}$ be Alice's encoder. We show that if $T'_A$ is given to Eve, then the scheme is as private as if Alice had sampled her encoder from $\mathcal{F}$. From the definition of the label configuration function, $\LC(T'_A\circ T_A) = C$ if and only if the $i$-th elements of $C$ is $C_i=L \circ T^{-1}_A \circ (T'_A)^{-1} (x_i)$. We define $\overline{x}_i = (T'_A)^{-1}(x_i)$, for every $x_i\in\mathcal{X}$, which is essentially a new ordering for $\mathcal{X}$. 
Under this new ordering, we denote the label configuration function by $\overline{\LC}(T_A)$. We note that $L \circ T^{-1}_A (\overline{x}_i) = L \circ T^{-1}_A \circ (T'_A)^{-1} (x_i) = C_i$, and therefore, $\overline{\LC}(T_A) = C$. We also define $\overline{Z} = (T'_A)^{-1}(Z)$, (Alice's encoded data under the new ordering). Thus,
\begin{multline*}
    \Pr[\bm{X_A} = X_A\mid \bm{T'_A\circ T_A ({X}_A)}=Z,\; \bm{\LC(T'_A\circ T_A})=C,\; \bm{T'_A}=T'] = \\ \Pr[\bm{{X}_A} = {X}_A\mid \bm{T_A ({X}_A)}=\overline{Z},\; \bm{\overline{\LC}(T_A})=C] ,
\end{multline*}
i.e., the same distribution as for $\mathcal{F}$ alone. Thus, $\mathcal{F}' \circ \mathcal{F}$ is at least as private as $\mathcal{F}$. Analogous arguments show that $\mathcal{F}' \circ \mathcal{F}$ is  also at least as private as $\mathcal{F}'$.

%% file: addit_experiments.tex

\textbf{Dataset licenses} Both the MIMIC-CXR and CheXpert datasets are publicly available under their own licenses. The MIMIC-CXR and CheXpert datasets are available under the PhysioNet Credentialed Health Data License 1.5.0 license and Stanford University School of Medicine CheXpert Dataset Research Use Agreement respectively.

\textbf{Computational cost} All experiments were conducted using Nvidia Tesla V100 or Nvidia RTX A6000 GPUs. All experiments took between 4-6 hours and were primarily bottlenecked by network bandwidth, as our images were hosted on an NFS server. \name models had approximately the same runtime as ViT.

\subsection{Evaluating modeling utility}

For each diagnosis task and training setting, we also report the AUC on MIMIC-CXR and CheXpert test set individually in Table  \ref{tab:xray-expers-mimic} and Table  \ref{tab:xray-expers-stanford} respectively.

\input{Tables/diagnosis_mimic}

\input{Tables/diagnosis_stanford}

\subsection{Additional attacks}

\textbf{Adversarial attacks.} 
We conducted additional MMD-based adversarial attacks on \name while varying the architecture of $T^*$. We wished to understand if an over-parameterized $T^*$, e.g with 2x the width or 3x the width, would have better success in attacking $T$. For these experiments, we used a hidden dimension of 2048 and a depth of 7 for $T$, and leveraged a $T^*$ with 2x and 3x the hidden dimension of $T$. We trained $T^*$ for 25 epochs using the Adam optimizer and an initial learning rate of $1e-04$. As shown in Table \ref{tab:attack-additional}, attacks leveraging an over-parameterized $T^*$ failed to uncover $T$, with a $\frac{T^* \text{MSE}}{T^\mu \text{MSE}} > 1$.

\input{Tables/addit_attacks}

\textbf{Plaintext attack.} While we do not assume that our adversary has access to either parallel data or the \name encoder in our threat model, we also investigated the robustness of \name to plaintext attacks. We hypothesized that given a sufficient amount of parallel data, i.e $(x_i, T(x_i))$ pairs and  $x_i\in \mathcal{X}$, an attacker could easily learn $T$. For this experiment, we leveraged the combined Mimic-CXR and CheXpert dataset with Cardiomegaly labels, used \name encoding with a depth of $7$ and a hidden dimension of $2048$, as in prior experiments. We trained an estimated $T^*$ for $50$ epochs using the Adam optimizer, an initial learning rate of $1\mathrm{e}{-04}$ and a batch size of $64$, and report $\frac{T^* \text{MSE}}{T^\mu \text{MSE}}$  on the test set.

We found that \name cannot defend against plaintext attacks, illustrating that the security of our encoding scheme relies on the lack of parallel data. We found that the plaintext attack obtains $\frac{T^* \text{MSE}}{T^\mu \text{MSE}}$ of $0.25 \pm 0.002$.

\subsection{Permutation encodings.} 
We hypothesized that for any \name encoder $T\in\mathcal{F}$ that maps $x_i$ to $T(x_i)$ for every $x_i\in X$ and a random permutation $\pi$, there exists a \name encoder $T_\pi \in \mathcal{F}$ that maps $x_i$ to $T(x_{\pi(i)})$ for every $x_i\in X$.
For this experiment, we used $128$ random samples from the combined Mimic-CXR and CheXpert dataset with cardiomegaly labels, and used a \name encoding with a depth of $7$ and hidden dimension of $2048$. We trained $T_{\pi}$ for $1000$ epochs to minimize $\text{MSE}(T(X), T_\pi (\pi(X)))$ using the Adam optimizer, an initial learning rate of $1\mathrm{e}{-04}$ and a batch size of $64$ and report $\frac{T^* \text{MSE}}{T^\mu \text{MSE}}$ on $128$ random samples. 

We found that $T_\pi$ could easily learn a mapping from $X$ to a randomly permuted $Z$. In this experiment, we found $T_\pi$ obtained an $\frac{T^* \text{MSE}}{T^\mu \text{MSE}}$ of $0.56 \pm 0.020$. We note that our permutation encoding experiment only included $128$ random samples, and it is unclear at what dataset size a fixed-size $T$ is no longer able to represent arbitrary $X$ to $Z$ bijections.

%% file: Tables/diagnosis_mimic.tex
\begin{table}[!htb]
\centering
\caption{Impact of \name on chest x-ray prediction tasks across different training settings. All metrics are ROC AUCs across the MIMIC-CXR test set. Combined-Clear and Combined-Private refer to using a single \name encoder across the combined MIMIC-CXR and CheXpert datasets and two independent \name encoders respectively. Guides of abbreviations for medical diagnosis: (E)dema, (P)neumothorax, (Co)nsolidation, (Ca)rdiomegaly and (A)telectasis.}
\label{tab:xray-expers-mimic}
\begin{tabular}{ccccccc}
\hline
\textit{Model} & E & P & Co & Ca & A & \textit{Average}\\ 
\hline
\multicolumn{7}{c}{\textbf{Train on MIMIC-CXR}}  \\
\hline
ViT & 88 $\pm $ 1  & 78 $\pm $ 3 & 77$ \pm $ 2 & 88 $\pm $ 1 & 85 $\pm $ 1 & \textit{83} \\
\hline
\name-ViT & 88 $\pm $ 2 & 81 $\pm $ 1 & 73 $\pm $ 2 & 88 $\pm $ 1 & 84 $\pm $ 1 & \textit{83} \\
\hline
\multicolumn{7}{c}{\textbf{Train on CheXpert}} \\
\hline
ViT & 80 $\pm $ 1 & 70 $\pm $ 1 & 69 $\pm $ 4 & 81 $\pm $ 1 & 77 $\pm $ 0 & \textit{75} \\
\hline
\name-ViT & 82 $\pm $ 1 & 69 $\pm $ 2 & 72 $\pm $ 3 & 79 $\pm $ 1 & 78 $\pm $ 0 & \textit{76} \\
\hline
\multicolumn{7}{c}{\textbf{Train on Combined-Clear}} \\
\hline
ViT & 89 $\pm $ 0 & 83 $\pm $ 1 & 77 $\pm $ 2 & 88 $\pm $ 0 & 86 $\pm $ 0 & \textit{84} \\
\hline
\name-ViT & 89 $\pm $ 0 & 82 $\pm $ 4 & 79 $\pm $ 1 & 88 $\pm $ 0 & 85 $\pm $ 0 & \textit{85} \\
\hline
\multicolumn{7}{c}{\textbf{Train on Combined-Private}} \\
\hline
\name-ViT & 90 $\pm $ 1 & 82$\pm $ 2 & 76 $\pm $ 3 & 88 $\pm $ 1 & 85 $\pm $ 1 & \textit{84}\\
\hline
\end{tabular}
\end{table}

%% file: Tables/diagnosis_stanford.tex
\begin{table}[!htb]
\centering
\caption{Impact of \name on chest x-ray prediction tasks across different training settings. All metrics are average ROC AUCs on CheXpert test set. Combined-Clear and Combined-Private refer to using a single \name encoder across the combined MIMIC-CXR and CheXpert datasets and two independent \name encoders respectively. Guides of abbreviations for medical diagnosis: (E)dema, (P)neumothorax, (Co)nsolidation, (Ca)rdiomegaly and (A)telectasis.}
\label{tab:xray-expers-stanford}
\begin{tabular}{ccccccc}
\hline
\textit{Model} & E & P & Co & Ca & A & \textit{Average}\\ 
\hline
\multicolumn{7}{c}{\textbf{Train on MIMIC-CXR}}  \\
\hline
ViT & 82 $\pm $ 1  & 61 $\pm $ 3 & 72$ \pm $ 2 & 86 $\pm $ 0 & 82 $\pm $ 1 & \textit{77} \\
\hline
\name-ViT & 83 $\pm $ 1 & 63 $\pm $ 2 & 71 $\pm $ 1 & 86 $\pm $ 0 & 83 $\pm $ 0 & \textit{77} \\
\hline
\multicolumn{7}{c}{\textbf{Train on CheXpert}} \\
\hline
ViT & 83 $\pm $ 1  & 72 $\pm $ 1 & 74$ \pm $ 1 & 85 $\pm $ 1 & 82 $\pm $ 0 & \textit{79} \\
\hline
\name-ViT & 85 $\pm $ 1 & 74 $\pm $ 0 & 78 $\pm $ 1 & 85 $\pm $ 1 & 83 $\pm $ 0 & \textit{81} \\
\hline
\multicolumn{7}{c}{\textbf{Train on Combined-Clear}} \\
\hline
ViT & 84 $\pm $ 1 & 71 $\pm $ 1 & 75 $\pm $ 2 & 86 $\pm $ 1 & 84 $\pm $ 1 & \textit{79} \\
\hline
\name-ViT & 85 $\pm $ 0 & 70 $\pm $ 2 & 78 $\pm $ 1 & 87 $\pm $ 0 & 84 $\pm $ 1 & \textit{81} \\
\hline
\multicolumn{7}{c}{\textbf{Train on Combined-Private}} \\
\hline
\name-ViT & 84 $\pm $ 2 & 71$\pm $ 3 & 77 $\pm $ 2 & 84 $\pm $ 2 & 82 $\pm $ 1 & \textit{80}\\
\hline
\end{tabular}
\end{table}


%% file: Tables/addit_attacks.tex
\begin{table}[!htb]
\centering
\caption{MSE of MMD-based adversarial attacks on \name encodings using an over-parametrized $T^*$.}
\label{tab:attack-additional}
\begin{tabular}{ c c}
\hline
Encoding &  $T^* \text{MSE} / T_{\mu} \text{MSE} $ \\ 
\hline
\name-width-2x  & 4.53 $\pm$ 0.05    \\
\hline
\name-width-3x  & 4.57 $\pm$ 0.08    \\
\hline
\end{tabular}

\end{table}